\documentclass[12pt]{article}
\usepackage{aasms4}                   
\usepackage{natbib}
\usepackage{graphicx}

\bibpunct{(}{)}{;}{a}{}{,}

\begin{document}                          

\title{Spectroscopic Signatures of Conduction-Mediated Transition Layers Above
an X-ray Illuminated Disk}   

\author{Yuexing Li, Ming F. Gu\altaffilmark{1,2}, and Steven M. Kahn}

\affil{Columbia Astrophysics Laboratory, Columbia University, New York,
NY 10027}
\affil{yxli@astro.columbia.edu, mfgu@space.mit.edu, skahn@astro.columbia.edu}
\altaffiltext{1}{Chandra Fellow}
\altaffiltext{2}{Now at: Center for Space Research, Massachusetts
Institute of Technology, Cambridge, MA 02139}
 
\begin{abstract}
We derive a semi-analytic solution for the structure of conduction-mediated
transition layers above an X-ray illuminated accretion disk, and calculate explicitly  
the X-ray line radiation resulting from both resonance line scattering and radiative
recombination in these layers. The vertical thermal structure of the
illuminated disk is found to depend on the illuminating continuum: for a hard
power law continuum, there are two stable phases connected by a single
transition layer; while for a softer continuum, there may exist three stable
phases connected by two separate transition layers, with an intermediate
stable layer in between. We show that the structure can be written as a
function of the electron scattering optical depth through these layers,  which
leads to unique predictions of the equivalent width of the resulting line
radiation from both recombination cascades and resonance line scattering. We
find that resonance line scattering plays an important role, especially for
the case where there is no intermediate stable layer. 

\end{abstract}

\keywords{accretion disks -- galaxies:nuclei -- radiative transfer -- resonant
scattering -- X-rays:spectra} 

\section{INTRODUCTION}
X-ray reflection off the surface of cold disks in active galactic
nuclei (AGN) and galactic black holes (GBHs) has been an active field of
research since the work of \citet{lightman88}. In early studies, the  illuminated
material was assumed to be cold and non-ionized
\citep{lightman88,george91}. It was soon realized, however, that
photoionization of the disk 
can have a great impact on both the reflected continuum and the iron fluorescence 
lines. Detailed calculations were then carried out by
\citet{ross93,matt93,zycki94,ross96} and \citet{ross99}. However, in all of these
papers, the density of the illuminated material was assumed to be constant
along the vertical direction. This assumption applies only to the simplest
version  of radiation-dominated Shakura-Sunyaev disks \citep{shakura73}, and only for
the portion where viscous dissipation is the dominating heating process. For the
surface layers, however, photoionization and Compton scattering are the
major heating sources. Therefore the approximation of constant density
is not appropriate. Moreover, thermal instability allows the coexistence of
 gas at different phases. These different phases have very different
temperatures, and hence different densities to keep the gas in pressure
balance.  

Recently \citet{nayakshin00} relaxed the simplifying assumption of constant gas
density. They determined the gas density from hydrostatic balance solved 
simultaneously with ionization balance and radiative transfer. They made an
important observation that the Thomson depth of the hot coronal layer can have
great influence on the X-ray reprocessing produced by the deeper, and much
cooler disk. In order to simplify the calculation of the vertical
structure, though, they ignored thermal conduction and the effects of
transition layers between the different stable phases. A discontinuous change
in temperature was allowed whenever an unstable 
phase was encountered. They argued that such transition layers are of little
importance because their Thomson depths are negligibly small. However,
without taking into account the role of thermal conduction, their method of
connecting two different stable layers is rather \textit{ad hoc}. Moreover,
even though the Thomson depths of these transition layers are small, it does
not guarantee that the X-ray emission and reflection from such layers are
negligible. Because the temperature regime where the transition layers exist
is not encountered in the stable phases, some of the most important lines can have appreciable
emissivity only in these layers. Also, since resonance line scattering has much
larger cross section than Thomson scattering, the optical depths in 
resonance lines can be significant. 
   
Including thermal conduction in the self-consistent solution of the vertical
structure presents a serious numerical challenge. The difficulties are
due to the coupling between hydrostatic balance, radiative transfer and
heat conduction. \citet{zeldovich69} first studied the phase 
equilibrium of a gas heated by cosmic rays and cooled
by radiation. They found that taking into account heat conduction in the
boundary layer allows one to obtain a unique solution of the stable
equilibrium. \citet{rozanska99} calculated the full temperature profile for a
Compton-heated corona, and \citet{rozanska00a} calculated the static conditions
of the plasma for different forms of heating and cooling. But
they did not include much discussion of the spectroscopic signatures resulting from
the derived vertical structure. 

In this paper, we first calculate the temperature structure in the
layers above the accretion disk, then calculate the emission lines via 
radiative recombination (RR) and reflection due to resonance line
scattering from the derived layers. Certain illuminating continua spectra
allow more than two stable phases to coexist, with two transition layers
connected by an intermediate stable layer. For the transition layer,
since the Thomson depth is small, the ionizing continuum can be treated as
constant; and since its geometric thickness is smaller than the pressure 
scale height, the pressure can be treated as constant as well. We can thus
obtain semi-analytic solution of the temperature profile by taking into
account thermal conduction.  For the intermediate stable layer, its
thickness is determined by the condition of hydrostatic equilibrium. In our
model, the normally incident continuum has a power-law spectrum 
with an energy index of $\alpha=1$. We also assume a plane-parallel
geometry and that the resonance line scattering is isotropic.

The structure of this paper is as follows: In \S\ref{sec_structure} we
discuss the existence of the thermal instability and compute
the thermal ionization structure of the transition layers; in
\S\ref{sec_spectrum} we calculate the recombination emission lines and the
reflection due to resonance line scattering; in \S\ref{sec_summary} we
summarize the important points of the calculations,  the validity of
various approximations made in the calculations, and the detectability of the
recombination emission and reflection lines.
 
\section{THERMAL INSTABILITY AND TRANSITION LAYERS}
\label{sec_structure}
\subsection{The ``S curves''}
\label{subsec_scurves}
The vertical structure of an X-ray illuminated disk at rest is governed by
the equations of hydrostatic equilibrium and  of energy conservation
\begin{eqnarray}
\label{eq_hydro}
\frac{dP}{dx} &=& -n_H(x)F, \\
\label{eq_transition}
-\frac{d}{dx}(\kappa(T)\frac{dT}{dx}) &=& \Phi(T, P, F_{x}(E)).
\end{eqnarray}
In the first equation, $F$ is the force density due to gravity and
radiation pressure. The dependence of the force on the plasma density is
included explicitly through the hydrogen density $n_H(x)$. In the second
equation, a time independent state is assumed, $\kappa(T)$ is the
thermal conductivity, and $\Phi(T, P, F_x(E))$ is the net heating rate
depending on the gas state and the incident flux $F_x(E)$ (differential
in energy).  We neglect the
effects of magnetic field and adopt the Spitzer conductivity appropriate for a
fully ionized plasma, 
$\kappa(T)=5.6\times 10^{-7} T^{5/2}$ erg cm$^2$ s$^{-1}$ K$^{-1}$
\citep{spitzer62}. We have used the classical heat flux,
$q_{class}=-\kappa(T)\nabla T$,  in Equation (\ref{eq_transition}) because the
electron mean free path is short compared to the temperature height scale.

Since the continuum flux may change along the
vertical height, in principle, the above two equations must be supplemented by
an equation for radiative transfer. A self-consistent solution of such equations is
difficult to obtain. In the following, we invoke a few physically
motivated approximations, which make the problem tractable.

\begin{figure}
\centerline{\includegraphics{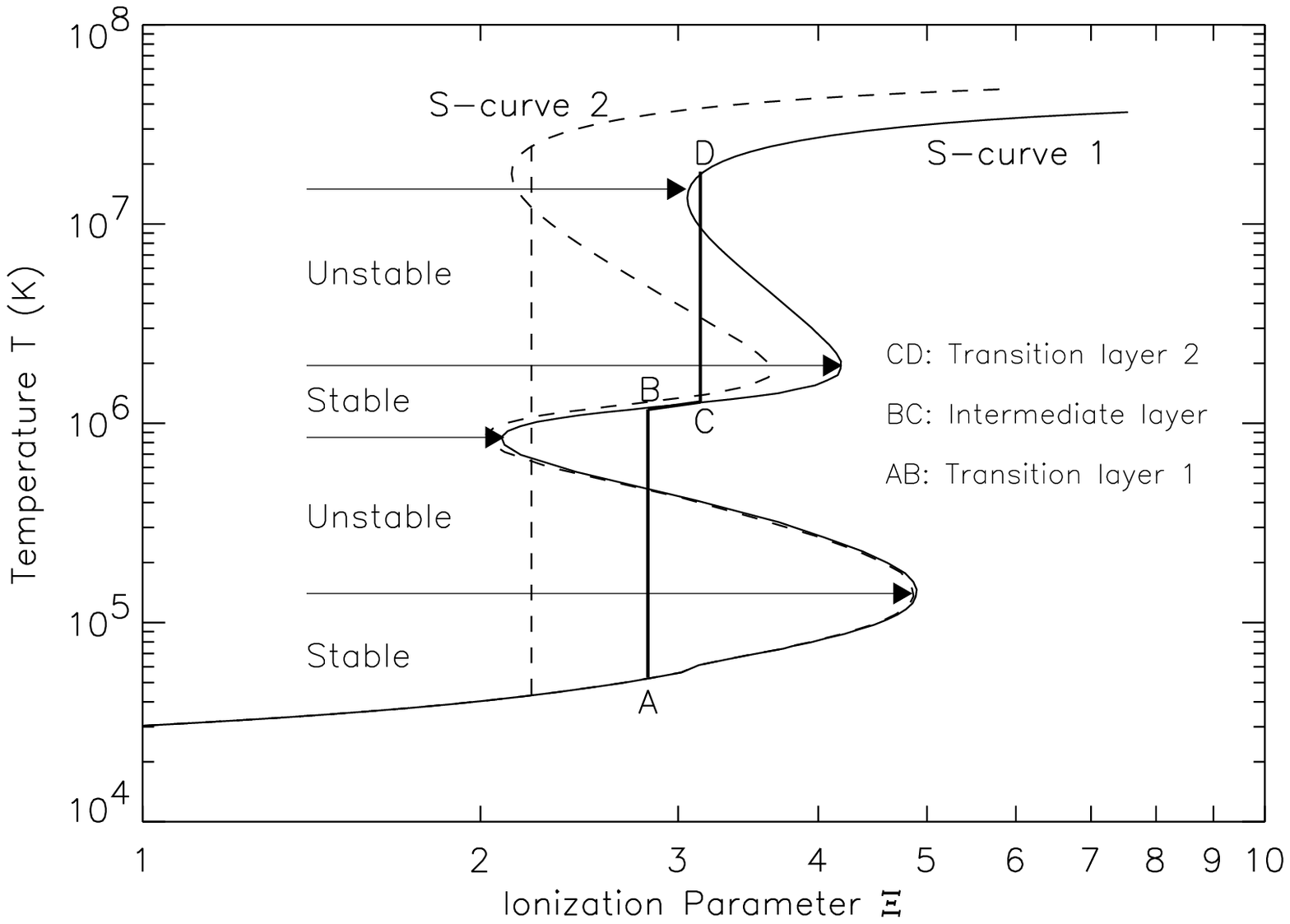}}
\caption{\label{fig_scurve}The ``S-curves'' produced by two different incident
ionizing spectra. The vertical line indicates the unique solution of $\Xi$
which satisfies Condition \ref{eq_y2}: two solutions for ``S-curve 1'',
$\Xi_1$ = 2.82 and $\Xi_2$ = 3.14; and only one for ``S-curve 2'', $\Xi$ = 2.22.} 
\end{figure}

First, in thermally stable regions, the gas temperature is slowly
varying, the heat conduction term in the energy balance equation can be
neglected. Therefore, the temperature can be determined locally with the
condition $\Phi(T, P, F_x(E)) = 0$. It is well known \citep{krolik81} that
the dependence of $\Phi$ on the gas pressure $P$ and the illuminating
continuum $F_x(E)$ can be expressed in the form of $\Phi = -n_e^2
\Lambda_{net}(T,\Xi)$, where $n_e$ is the electron density,
$\Lambda_{net}(T,\Xi)$ is the net cooling rate per unit volume, and $\Xi$ is
an ionization parameter defined by  
\begin{equation}
\Xi=\frac{F_x}{cP}, 
\end{equation}
where $F_x$ is the total flux of the continuum, and $c$ is the speed of light. In
Figure \ref{fig_scurve}, we show the local energy equilibrium curve $T$ versus
$\Xi$,  at $\Lambda_{net}(T, \Xi) = 0$ calculated with the photoionization
code XSTAR \citep{kallman86}. These curves are commonly referred to as ``S
curves'' due to their appearance. The illuminating continuum is assumed to be a
power law with energy index $\alpha=1$. The solid line labeled with ``S-curve
1'' corresponds to a low energy cutoff at 1 eV and a high energy cutoff at 150
keV, while the dashed line ``S-curve 2'' corresponds to a high energy cutoff
at 200 keV. The choice of such incident spectra is based on their common
appearance in many AGNs and GBHCs. The region is thermally ``unstable'' where
the ``S-curve'' has a negative slope,  and ``stable'' where the slope is
positive, as indicated in Figure \ref{fig_scurve}. In the thermally unstable regions, we have 
\begin{equation}
\label{eq_instability}
\left(\frac{\partial \Xi}{\partial T}\right)_{\Lambda_{net}=0} < 0,
\end{equation}
where the derivative is taken while the energy balance is satisfied, i.e.,
$\Lambda_{net}=0$. This condition was shown \citep{krolik81} to be equivalent
to the instability condition discovered by \citet{field65}
\begin{equation}
\left(\frac{\partial \Lambda_{net}}{\partial T}\right)_{P} < 0.
\end{equation}

\placefigure{fig_scurve}

\subsection{The Temperature Profiles}
\label{subsec_tprofile}
\subsubsection{The Transition layers}
The thermal instability allows the gas to coexist at different phases. The gas
temperature may change by orders of magnitude over a geometric thin 
region whenever an unstable phase separates two stable ones. This results in
enormous temperature gradients and heat conduction. Therefore the heat
conduction in the energy balance equation should be included in such transition
layers between stable phases. On the other hand, the thicknesses of these transition
layers are usually smaller than the pressure scale height, so one can safely
treat the gas pressure as constant in these regions. Moreover, the
continuum radiative transfer can  be neglected because the Compton optical
depth is found to be small. The vertical structure of the transition regions
is then solely determined by the energy balance equation with heat
conduction. The Thomson optical depth $d\tau_e = n_e\sigma_T dx$ of such
regions is readily estimated by 
\begin{equation}
\Delta \tau_e = \frac{\kappa(T) T \sigma_T}{\sqrt{\Lambda_{net}(T, \Xi)}}.
\end{equation}
where $\sigma_T$ is the Thomson scattering cross section. 

The transition layer solution is not arbitrary under the steady-state
conditions, i.e., where there is no mass exchange between the two stable
phases which the transition layer connects. A similar problem in the context
of interstellar gas heated by cosmic rays was well studied by
\citet{zeldovich69}. We follow their procedure here and define
$y=\kappa(T)dT/d\tau_e$, in order to rewrite Equation 
(\ref{eq_transition}) in the form: 
\begin{equation}
\label{eq_dy2}
\frac{dy^{2}}{dT}=-\frac{2\kappa(T)\Lambda_{net}(T,\Xi)}{\sigma_T^2}.
\end{equation}
A steady-state requires vanishing heat flux at both boundaries of the
transition layer, or
\begin{equation}
\label{eq_y2}
y^2\mid_{T=T_1}-y^2\mid_{T=T_2}=0=-2\int_{T_1}^{T_2}
\frac{\kappa(T)\Lambda_{net}(T,\Xi)}{\sigma_T^2}dT, 
\end{equation}
where $T_1$ and $T_2$ are the temperatures of the two stable phases which are
connected by a transition layer. This condition determines a unique ionization
parameter $\Xi$ for the transition layer in question, and the integration of
Equation (\ref{eq_dy2}) along the vertical height gives the detailed temperature
profile as a function of optical depth. 

If the disk does not realize the steady-state solution, there are additional
enthalpy terms in Equation (\ref{eq_transition}) which require that there be mass
flow through the transition region -- i.e. the cool material in the disk  
evaporates, or the hot material in the corona  condenses
\citep{meyer94,dullemond99,rozanska00b}. Physically, this
corresponds to a movement of the transition layer up or down through the
vertical disk structure. However, since the density increases monotonically
toward the center of the disk, this ``motion''  stops where the transition
layer reaches the steady state value of $\Xi$. Thus, in the absence of disk
winds or continuous condensation from a disk corona, the steady state solution
should generally be obtained. 

\subsubsection{The Intermediate Stable Layer}
There is a complication for the ``S-curves'' shown in Figure \ref{fig_scurve},
because in each curve there exist two unstable regions and therefore there
should be two transition layers.
For ``S curve 1'', Condition \ref{eq_y2} can be met for both transition layers
with resulting ionization parameters $\Xi_1 < \Xi_2$, where $\Xi_1$ and
$\Xi_2$ are associated with the transition layer which connects to the lowest
temperature phase and highest temperature phase, respectively. For ``S curve
2'', the resulting ionization parameters for two transition layers, however,
satisfy $\Xi_1 > \Xi_2$. Such a situation is unphysical, where the ionization
parameter of the upper transition layer is smaller than that of the
lower one, because in the context of accretion disks, the upper layer receives
more ionizing flux and has lower pressure. So in
practice, for ``S curve 2'' the intermediate stable region is skipped and a
transition layer connects the lowest temperature phase to the highest
temperature phase directly. The ionization parameter of this transition layer
is determined the same way by applying Equation (\ref{eq_y2}).  

There is then an intermediate stable layer, of nearly uniform temperature, in
between these two transition layers for ``S curve 1'', as indicated with BC in
Figure \ref{fig_scurve}. The thickness of the intermediate layer should in
principle be obtained by solving the coupled equations of hydrostatic
equilibrium, energy balance, and radiative transfer. However, unlike the 
stable phase at the disk base or that of the corona, which may be Compton
thick, this intermediate stable layer is generally optically thin, because
its optical depth is restricted by the difference between $\Xi_1$ and
$\Xi_2$. Furthermore, the temperature in this layer is slowly varying, and
therefore heat conduction can be neglected. We shall make another
approximation that the variation of the force density $F$ in the hydrostatic
equation can also be neglected. This may not be a good
approximation. However, since our main purpose is to investigate qualitatively the
effects of an intermediate stable layer, such a simplifying procedure does
capture the proper scaling relations of the problem, and has the advantage of
less specific model dependence. Writing Equation (\ref{eq_hydro}) in a
dimensionless form and parameterizing the force factor by a dimensionless
parameter $A=cF/\sigma_TF_x$, we obtain 
\begin{equation}
\label{eq_hydroapp}
\frac{1}{\Xi^2}\frac{d\Xi}{d\tau_e}=A.
\end{equation}
The parameter A defined here is identical to the gravity parameter in
\citet{nayakshin00} in the absence of radiation pressure. The
integration of this 
equation from $\Xi_1$ to $\Xi_2$ gives the Thomson depth of
the intermediate stable layer. 

Assuming  radiation pressure can be neglected, The force factor at a given
radius $R$ of the disk can be estimated as 
\begin{equation}
A=\frac{1}{l}\frac{H}{R}\frac{4\pi GMm_pc}{\sigma_TL_{EDD}},
\end{equation}
where $l$ is the luminosity of the continuum source in  units of the Eddington
limit, $H$ is the half thickness of the disk at radius $R$, $M$ is the mass of
the central source, $m_p$ is the proton mass, and $L_{EDD}$ is the Eddington
luminosity. In this estimate, we have assumed that $F_x=L/4\pi R^2$. This
gives $A$ the value:
\begin{equation}
A = 1.4 \frac{1}{l}\frac{H}{R}.
\end{equation}
For a typical thin disk, as those present in AGN and black hole binaries, one
expects $H/R \sim 0.01$. If the luminosity is sub Eddington, $l \sim 0.01$ as
in most AGNs, $A$ is of order unity. However, since the disk surface may not
be normal to the continuum radiation, $F_x$ may be only a fraction of the
value assumed above, which increases $A$ by one or two orders of magnitude. On
the other hand, as the source approaches the Eddington limit, $A$ may become
smaller than 1. Therefore, we expect $A$ to be in the range of 0.1 -- 10. The
exceptional cases of much smaller and larger $A$ are discussed in
\S\ref{sec_summary}.  

\begin{figure}
\centerline{\includegraphics{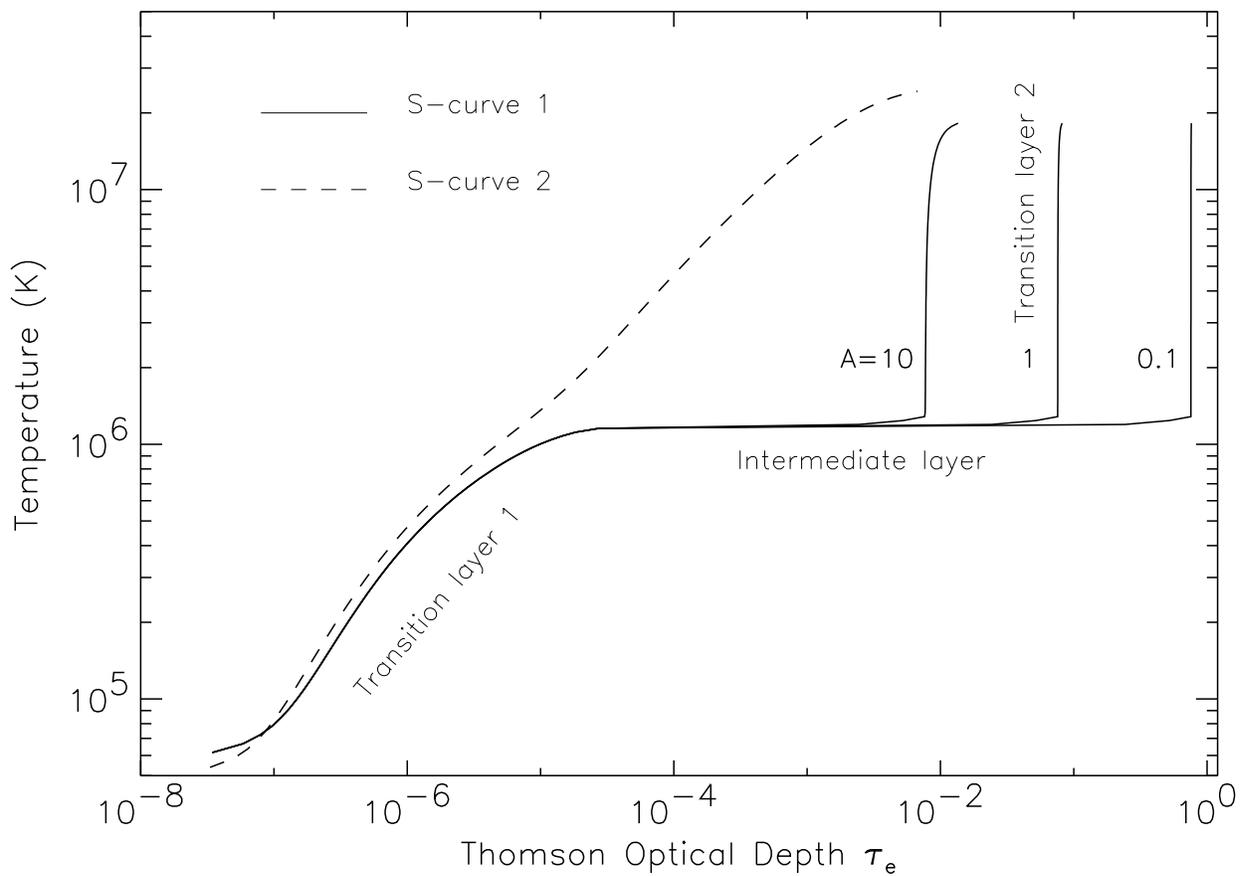}}
\caption{\label{fig_transition}The temperature profiles of the transition
layers and the intermediate stable layer versus Thomson optical depth $\tau_e$. The
solid curves correspond to ``S-curve 1'' with different force factors $A$, the
dashed line corresponds to the ``S-curve 2''. }
\end{figure}

The temperature profiles and optical depths of the transition layers and
possible intermediate 
stable layer are shown in Figures \ref{fig_transition}. Three labeled curves
in solid lines correspond to ``S-curve 1'' (in Figure \ref{fig_scurve}) with 
different values of the force factors $A$ = 10, 1, and 0.1, respectively.  The
dashed curve corresponds to  ``S-curve 2'', where there is no
intermediate stable layer. For each of the solid curves, three layers are
clearly seen, with two transition layers being connected by an intermediate
stable layer, as illustrated for the case of $A=1$. The smaller $A$ produces a
more extended intermediate stable layer as expected.

\placefigure{fig_transition}
 
\section{X-RAY EMISSION AND RESONANCE LINE SCATTERING}
\label{sec_spectrum}
\subsection{X-ray Emission From Radiative Recombination (RR)}
\label{subsec_emission}
Because the stable phase with the lowest temperature is almost neutral, and
the stable phase at the highest temperature is almost fully ionized, they are
not efficient 
in generating X-ray line emission, except for iron fluorescence lines from the
neutral material. Only the transition layers and the intermediate stable layer
are expected to emit discrete lines in the soft X-ray band. In a photoionized
plasma, the temperature is too low for collisional excitation to be an
important line formation process. Instead, radiative recombination (RR)
followed by cascades dominates the line emission. The flux of a particular line
can be written as
\begin{equation}
\label{flux_emission}
F_{em}=\int n_{e}n_{i}E_l\Lambda_{l}(T)dx,
\end{equation}
where $n_i$ is the density of the ion before recombination, $E_l$ is the line
energy, and $\Lambda_{l}(T)$ is the line emissivity defined as in \citet{sako99}. 

In ionization equilibrium where the ionization  rate is equal to the
recombination rate,  we have
$n_{e}n_{i} \alpha_{i} = n_{i-1} \int \frac{F_x(E)}{E} \sigma_{i-1}(E) dE$, 
where $\alpha_{i}$ is the recombination coefficient of ion $i$, $n_{i-1}$ is the
number density of ion $i-1$,  $F_x(E)$ is again the monochromatic incident flux
(differential in energy), and $\sigma_{i-1}(E)$ is the photoionization cross
section of ion $i-1$. Defining the branching ratio $\rho_l(T) =
\frac{\Lambda_{l}}{\alpha_i}$, Equation (\ref{flux_emission}) can be rewritten as:
\begin{equation}
\label{f_em}
F_{em}=E_l\int\frac{F_x(E)}{E}\frac{\sigma_{i-1}(E)}{\sigma_T}dE\int\rho_l(T)f_{i-1}(\Xi,T)d\tau_e,   
\end{equation}
where $f_{i-1}(T,\Xi) = \frac{n_{i-1}}{n_e}$ is the fractional abundance of
ion $i-1$ with respect to the electron density.

As indicated, $\rho_l$ and $f_{i-1}$ depend only on temperature $T$ and
ionization parameter $\Xi$, which are both functions of optical depth 
$\tau_e$. For convenience, we further define the ``emission equivalent width''
$EW_{l}^{e} = \frac{F_{em}}{F_x(E_l)}$. Then if $h(E) =
\frac{F_x(E)}{F_x(E_l)}$, which depends only on the shape of the incident
continuum, $EW_{l}^{e}$  can be written as:
\begin{equation}
\label{em_ew}
EW_{l}^{e}=E_l\int\frac{h(E)}{E}\frac{\sigma_{i-1}(E)}{\sigma_T}dE\int\rho_l(T)f_{i-1}(T,\Xi)d\tau_e.  
\end{equation}

Therefore the emission equivalent width $EW_{l}^{e}$ is independent of the
density of the medium and the incident flux. It is a unique function of the
structure deduced in Section \ref{sec_structure}. All other variables depend
only on $T$ and $\Xi$. The numerical values of $\rho_l(T)$ were
provided by D. Liedahl (private communication), and were calculated using the
models described in \citet{sako99}. The values of $f_{i-1}(T,\Xi)$ were
computed using XSTAR \citep{kallman86}.

\begin{figure}
\centerline{\includegraphics{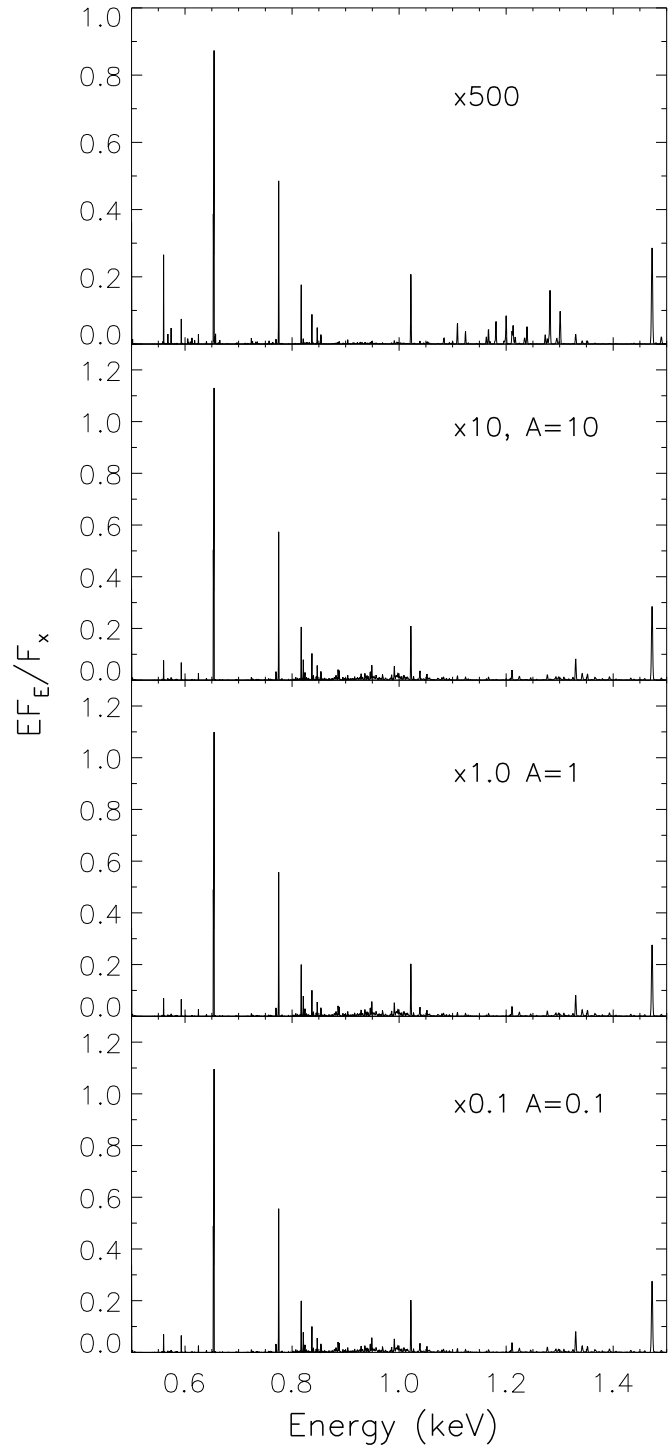}}
\caption{\label{fig_emission}The spectra of the emission lines via RR 
in the transition layers and the intermediate stable layer. The spectral
resolution is $\Delta E \sim 2$ eV. The top panel corresponds to 
``S-curve 2'', while bottom panels correspond to ``S-curve 1'' with different
parameter A. } 
\end{figure}

In Figure \ref{fig_emission}, we plot the spectra of the recombination
emission within the 0.5 -- 1.5 keV band with 
a spectral resolution $\sim 2$ eV,  which is close to the spectral
resolution of the grating spectrometers on \textit{Chandra} and
\textit{XMM-Newton}. The top panel corresponds to the case 
without an intermediate stable layer, and the bottom panels corresponds to the
case with an intermediate layer for $A=10, 1$ and 0.1, respectively. For clarity,
from top to bottom, the flux in each panel is multiplied by a factor indicated
in each panel.  It appears that the existence of an
intermediate stable layer enhances the emission in this energy band. This is
not surprising since the ions that are responsible for these lines have peak
abundances at temperatures close to that of the intermediate stable layer. In
all cases, the equivalent widths (EWs) of the emission lines are less than 1.0
eV, with respect to the ionizing continuum. The strongest lines are the
hydrogen-like and helium-like lines of oxygen, with EWs approaching several tenths
of an eV. Hydrogen-like and helium-like lines from iron outside the plotted
energy band are somewhat stronger, with EWs reaching a few eV. We note that
our low equivalent width values conflicts with those derived by
\citet{nayakshin00}, who found some lines with equivalent widths as high as 30
eV. However, since they did not consider the appropriate locations for the
transition regions in $\Xi$-space, their intermediate stable layer subtended a
much larger optical depth than we find here. Naturally with a thicker layer,
they found larger equivalent widths. 

\placefigure{fig_emission}

\subsection{X-ray Reflection From Resonance Scattering (RS)}
\label{subsec_reflection}
Emission from RR is not the only line formation process in the
 transition layers and the intermediate stable layer. Due to very large
cross sections in resonance line scattering, the reflected flux in these lines
may be significant. With the computed thermal and ionization structure, the
column density in each ion and the optical depth in all resonance lines can be
calculated straightforwardly. 

The cross sections for resonance line scattering depend on the line broadening. We
assume thermal Doppler effects as the only mechanism. Although the gas
temperature is a function of depth, we calculate the line width for a temperature
where the abundance of each ion peaks as an average, and assume that the resonance
scattering cross sections are uniform along the vertical direction. In terms
of absorption oscillator strength $f$, this cross section of the resonance
line scattering $\sigma_{RS}$ can be written as
\begin{equation}
\sigma_{RS} = \frac{\pi e^2}{m_ec^2}f\frac{\lambda^2}{\Delta\lambda},
\end{equation}
where $m_e$ is the electron mass, $e$ is the electron charge, $\lambda$ is the
wavelength of the line and $\Delta\lambda$ is the average line width in
wavelength. Under the assumption of thermal Doppler broadening,
\begin{equation}
\Delta\lambda=2.35\lambda\sqrt{\frac{kT_M}{m_ic^2}},
\end{equation}
where $T_M$ is the temperature at which the ion abundance peaks, and $m_i$ is
the ion mass. The resonance scattering optical depth $\tau_{RS}$ for a line
from ion $i$ can be estimated as 
\begin{equation}
\tau_{RS}=\sigma_{RS} N_i,
\end{equation}
where $N_i$ is the column density of the ion. 

\begin{figure}
\centerline{\includegraphics{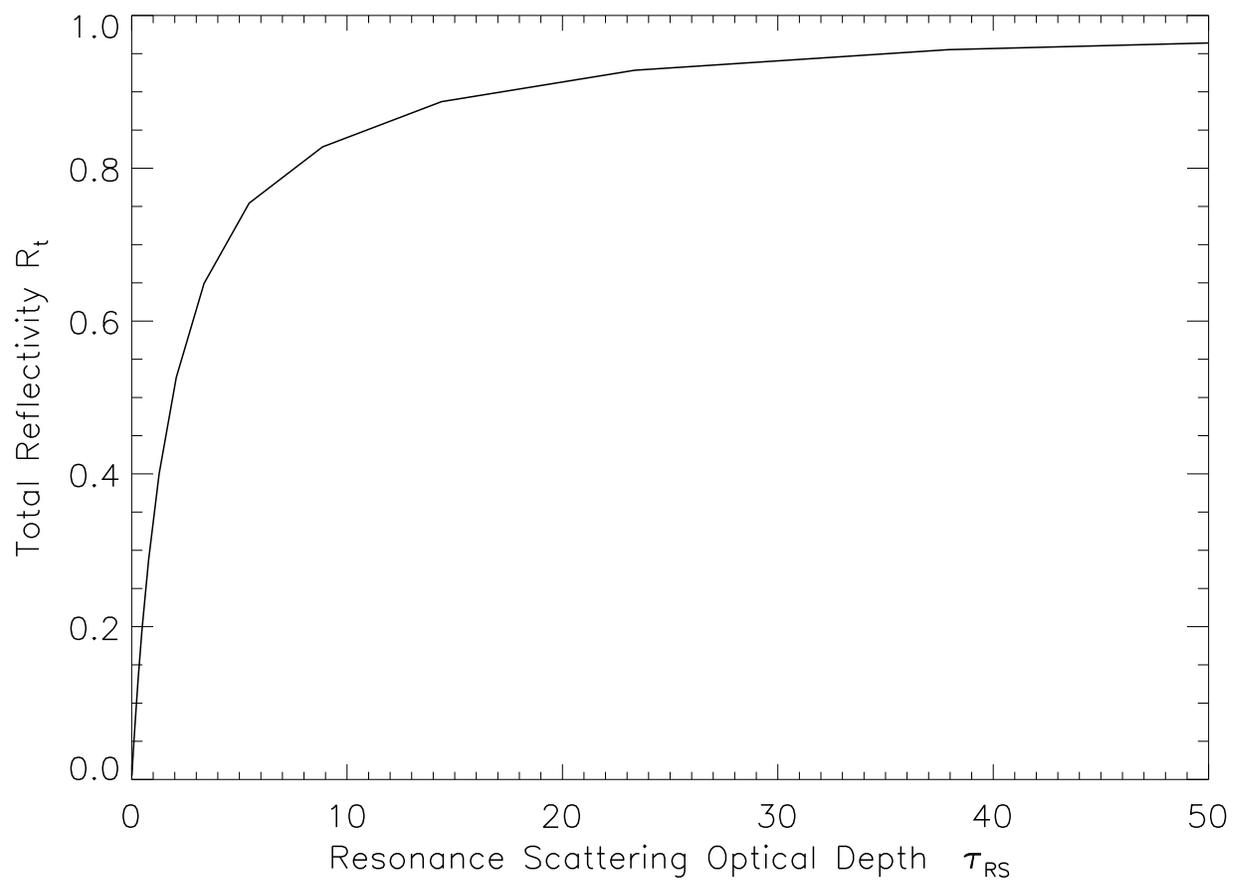}}
\caption{\label{fig_totref}The angle integrated reflectivity of a plane
parallel slab with normal incident flux as a function of the resonance
scattering optical depth $\tau_{RS}$.}
\end{figure}

The radiative transfer in the line is a complicated issue \citep{dumont00}. A
full treatment is beyond the scope of this work. However, since we are only
interested in a reasonable estimate of the reflected line flux, a simple
approach may be adopted. We assume the resonance line scattering is isotropic and
conservative and neglect the polarization dependence. Under such
conditions, the reflection and transmission contributions by a plane-parallel slab of finite
optical depth $\tau$ have been solved by \citet{chandra60}. For normal incident
flux $F_x$, the angle dependent reflectivity is 
\begin{equation} 
r(\mu)=\frac{I}{F_x}=\frac{S(\tau,\mu)}{4\pi\mu},
\end{equation} 
where $\mu=\cos\theta$, $r(\mu)$ is the reflectivity at $\mu$, $I$ is the 
reflected intensity, $F_x$ is the incident flux, and $S(\tau,\mu)$ is the 
scattering function defined as
\begin{equation} 
(1+\frac{1}{\mu})S(\tau,\mu)=X(\mu)X(1)-Y(\mu)Y(1),
\end{equation} 
where $X(\mu)$ and $Y(\mu)$ are two functions that satisfy the following
integral equations:
\begin{eqnarray} 
X(\mu)&=& 1+\frac{\mu}{2}\int_{0}^{1}\frac{1}{\mu+\mu^{\prime}}
[X(\mu)X(\mu^{\prime})-Y(\mu)Y(\mu^{\prime})]d\mu^{\prime} \nonumber \\
Y(\mu)&=& e^{-\tau /\mu}+\frac{\mu}{2}\int_{0}^{1}\frac{1}{\mu-\mu^{\prime}}[Y(\mu)X(\mu^{\prime})-X(\mu)Y(\mu^{\prime})]d\mu^{\prime}.
\end{eqnarray}
The solutions of these equations may be obtained by an iterative method with
the starting point $X(\mu)=1$ and $Y(\mu)=e^{-\tau/\mu}$. The angle integrated
reflectivity $R_t$ can be calculated as
\begin{equation}
R_t=2\pi\int_{0}^{1}r(\mu)\mu d\mu,
\end{equation}    
and is shown in Figure \ref{fig_totref} as a function of the resonance
scattering optical depth $\tau_{RS}$.

\placefigure{fig_totref}

The reflected flux in a line can be written as
\begin{equation}
F_r=R_{t}F_x(E)\Delta E,
\end{equation}
where $\Delta E$ is the line width in energy. Similarly to the ``emission
equivalent width'' $EW_{l}^{e}$, we define a ``reflection equivalent width''
$EW_{l}^{r} = \frac{F_r}{F_x(E_l)}$, which results in:
\begin{equation} 
\label{rf_ew}
EW_{l}^{r}=\frac{R_t\Delta\lambda hc}{\lambda^{2}}.
\end{equation}

This ``reflection equivalent width''
from our numerical results is a few tenths of an eV for strong resonance
lines, similar to that of the recombination emission lines. 
 
\begin{figure}
\centerline{\includegraphics{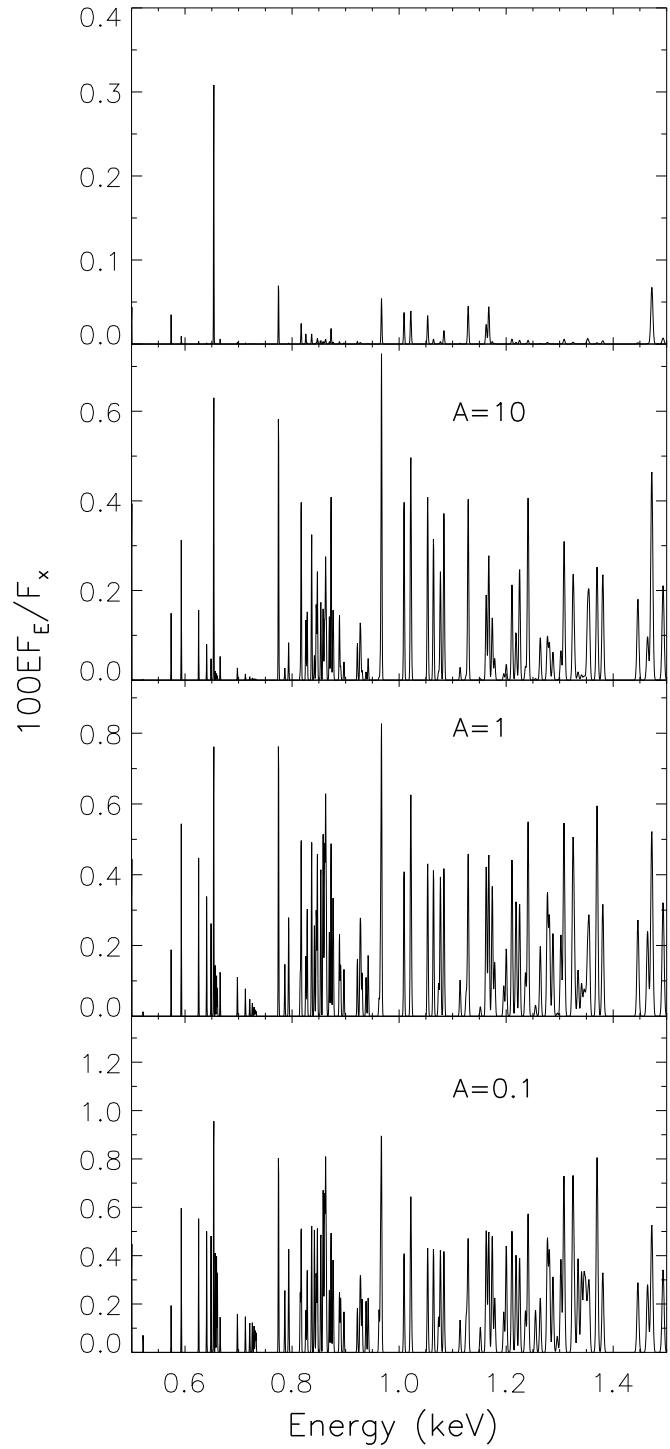}}
\caption{\label{fig_reflection}The spectra of the reflection lines due to resonance
scattering in the transition layers and the intermediate stable layer. The
spectral resolution is $\sim 2$ eV. The top panel corresponds to 
``S-curve 2'', while bottom panels correspond to ``S-curve 1'' with different
parameter A.} 
\end{figure}

In Figure \ref{fig_reflection}, we plot the spectra of the resonantly scattered 
lines in the energy band 0.5 -- 1.5 keV with a spectral resolution $\sim 2$
eV. The top panel corresponds to the case  
without an intermediate stable layer, and the bottom panels corresponds to the
case with an intermediate layer for $A=10, 1$ and 0.1, respectively. For clarity,
from top to bottom, the flux in each panel is multiplied by a factor of
100. We see that the equivalent widths of the reflected lines are notably
enhanced when there is an intermediate stable layer, but not as significantly
as for the recombination emission lines. This is because the optical depths of
many strong lines become much larger than unity, and the reflection is
saturated. If there are broadening mechanisms other than thermal Doppler
effects, such as turbulent velocity, the reflected intensity can be further
enhanced. 

\placefigure{fig_reflection}

\subsection{X-ray Spectrum with RR Emission and RS Reflection} 
\begin{figure}
\centerline{\includegraphics{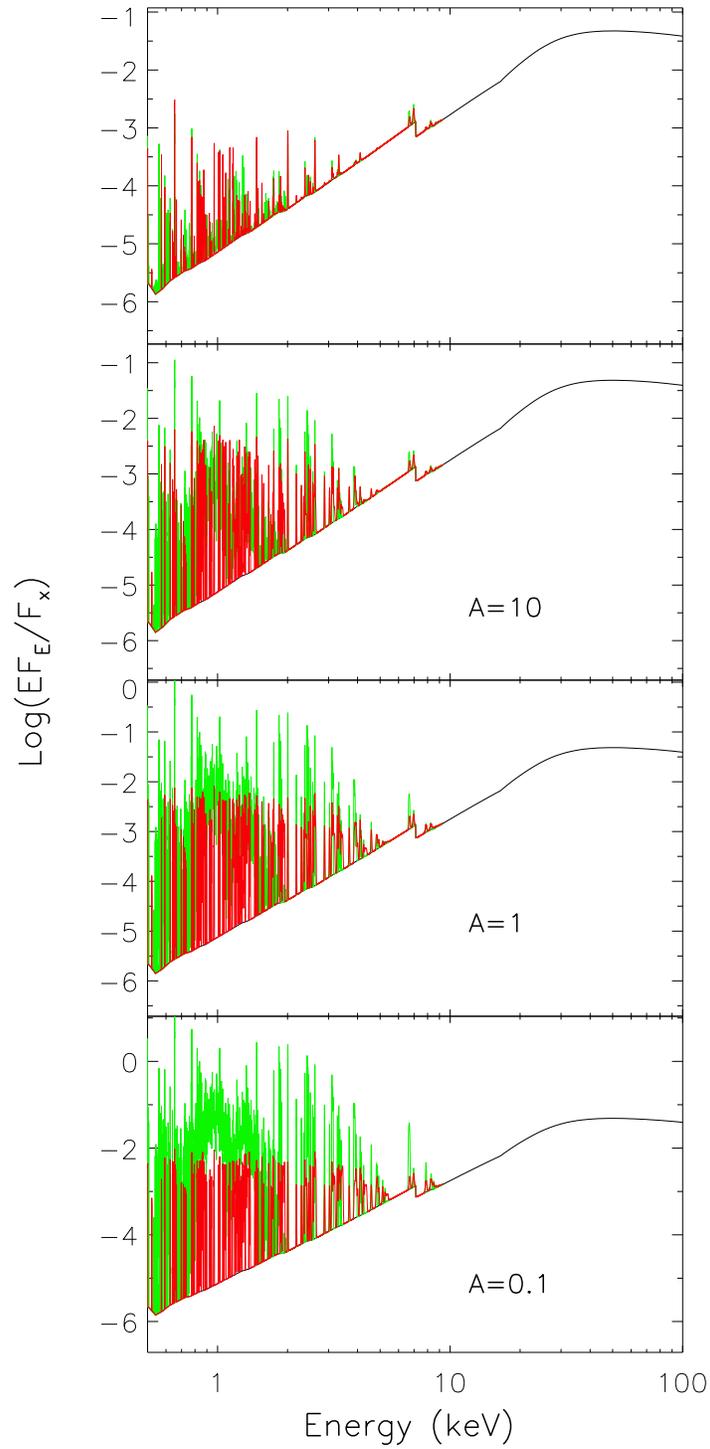}}
\caption{\label{fig_spectrum}The combined spectrum. Green -- emission lines
via recombination, red -- reflection lines due to resonance line scattering, and
black -- the Compton reflection hump. The spectral resolution is $ \sim 2$
eV. The top panel corresponds to ``S-curve 2'', while bottom panels correspond
to ``S-curve 1'' with different parameter A.} 
\end{figure}

In order to gain a crude idea of the relative importance of recombination emission and
reflection from the transition layers and intermediate stable layer, we
compare them to the ``hump'' produced by Compton scattering off a cold surface
\citep{lightman88}. We use the Greens function obtained by \citet{lightman88}
to calculate the Compton reflection. This method was verified to be accurate
with a Monte Carlo procedure by \citet{george91}. In Figure
\ref{fig_spectrum}, we show the combined spectra including recombination emission and
reflection lines from resonance line scattering, and the Compton reflection
``hump''.   
 
\placefigure{fig_spectrum}

\section{SUMMARY}
\label{sec_summary}
We now summarize the most important conclusions that can be drawn from the
calculations presented in this paper. We also discuss the detectability of the
predicted line features.  
\begin{enumerate}
\item The unique ionization parameters that characterize the steady-state
solutions of the transition layers depend on the shape of the ``S-curve''. We
have shown that two power-law illuminating spectra with different high energy
cutoffs produce very different temperature profiles. The harder spectrum only
allows one transition layer even though there are two unstable branches in the
``S-curve'', while the softer one allows two separate transition layers  
connected by an intermediate stable layer. This is due to the fact that the
ionization parameter of the upper transition layer must be larger than that of
the lower one, if they are to exist separately in a disk environment. The
harder spectrum produces a turnover point of the upper branch of the
``S-curve'' at smaller $\Xi$. Therefore the transition layer due to the upper 
unstable region joins the lower one smoothly without allowing the intermediate stable
region to form. The turnover of the upper ``S-curve'' represents the point
where Compton heating starts to overwhelm bremsstrahlung. The ionization
parameter at which this point occurs is related to the Compton temperature of
the continuum, $\Xi_{top} \propto T_{IC}^{-3/2}$ 
\citep{krolik81, nayakshin00a}. A harder spectrum has larger $T_{IC}$,
therefore the intermediate stable layer tends to disappear for hard incident
spectra. 

\item Although the Thomson depths of the transition layers and possible
intermediate stable layer are generally negligible, 
the X-ray emission lines from them may comprise the main observable line
features, because the temperatures of these layers are inaccessible to the
stable phases, and thus some of the important lines can have
appreciable emissivity only in these layers. Due to the much larger cross
sections for resonance line scattering, reflection due to resonance lines off such
transition layers is also important. The strengths of reflected lines are at
least comparable with those of the recombination emission lines when there is no
intermediate stable layer. Because the appearance of the reflected line
spectrum is different from that of the recombination emission spectrum, high
resolution spectroscopic observations should be able to distinguish these
mechanisms.  

\item The justification of the assumption that the ionizing continuum does not
scatter in the intermediate layer depends on the magnitude of the parameter 
$A$. The Thomson depth of this layer $\tau_e$ is given by: 
\begin{equation}
\tau_e=\frac{1}{A}(\frac{1}{\Xi_1}-\frac{1}{\Xi_2}).
\end{equation}
For the power-law continuum with high energy cutoff at 150 keV (``S curve
1''), $\Xi_1=2.82$ and $\Xi_2=3.14$ from our numerical results. Therefore,
$\tau_e=0.036/A$. $\tau_e$ is much less than unity as long as $A$ is greater
than 0.1. For smaller $A$, however, another 
effect comes into play. \citet{nayakshin00} showed that the Thomson depth of
the coronal layer (the stable phase with highest temperature) exceeds unity
when the gravity parameter (identical to $A$ defined here when the radiation
pressure is neglected) is $\sim$
0.01. Therefore not much ionizing flux can penetrate this layer, and the
reprocessing in the deeper and cooler layers can be neglected completely. As
$A$ becomes much larger than 10, the thickness of the intermediate stable
layer is negligible compared to the transition layers. Therefore, its presence
may be ignored. 

\item Since the recombination rate must equal the photoionization rate in the
irradiated gas, recombination radiation is also a form of reflection --
i.e. the line equivalent widths are independent of the incident flux. They
depend only on the structure ($T(\tau); \Xi(\tau)$) deduced from the
hydrostatic and energy balance equations.

\item The detectability of these recombination emission and reflection lines depends on
whether the primary continuum is viewed directly. When the ionizing continuum is in 
direct view, our results show that the EWs of the strongest lines in the 0.5 
-- 1.5 keV band are at most a few tenths of an eV, slightly larger when there is an
intermediate stable layer. 

The signal to noise ratio (SNR) in such a line can be written as  
\begin{equation}
SNR=\frac{EW}{E_l}(tE_lF_{\gamma})^{1/2}\left(A_{eff} \frac{E}{\delta E}\right)^{1/2},
\end{equation}
where $t$ is the integration time of the observation, $E_l$ is the energy of
the line, $F_{\gamma}$ is the photon flux
in the continuum,  $A_{eff}$ and $E/\delta E$ are the
effective area and resolving power of the instrument, respectively. For a line
at $\sim$ 1 keV, with $EW=1 eV$, and with HETGS on board \textit{Chandra}, we have
$SNR=0.25\sqrt{tF_{\gamma}}$. A typical Seyfert 1 galaxy has a  flux of
$10^{-10}$ erg cm$^{-2}$ s$^{-1}$ in the energy band of 2--10 keV. Assuming a
power law with energy index of 1, 
the photon flux at 1 keV would be $\sim 0.05$ cm$^{-2}$ s$^{-1}$ keV$^{-1}$.
For a reasonable integration time of 10 ks, we have SNR $\sim 5$. When the
primary continuum is obscured as in Seyfert 2 galaxies, the EWs of the
emission and reflection lines can be orders of magnitude larger, because the
continuum at this energy region is absorbed severely, and the SNR can be
greatly enhanced, making these lines observable. 
\end{enumerate}

Acknowledgment:
SMK acknowledges several grants from NASA which partially supported this work,
MFG acknowledges the support of a Chandra Fellowship at MIT. We wish to thank
M. Sako, D. Savin and E. Behar for several useful discussions.

\clearpage

\end{document}